\documentclass{article}                

\usepackage{opex3}
\usepackage{here}
\usepackage{graphicx}

\newcommand{\be}{\begin{equation}}
\newcommand{\ee}{\end{equation}}

\newcommand\rpict[1]{\ref{fig:#1}}

\newcommand\leqt[1]{\protect\label{eq:#1}}

\newcounter{Fig}

\begin{document}
\begin{sloppy}
\title{Discrete interband mutual focusing in nonlinear photonic lattices}

\author{Christian R. Rosberg$^{1,2,3}$, Brendan Hanna$^{2,3}$, Dragomir N. Neshev$^{1}$,
Andrey A. Sukhorukov$^{1,3}$, Wieslaw Krolikowski$^{2,3}$, \\and
Yuri S. Kivshar$^{1,3}$}

\address{$^{1}$Nonlinear Physics Centre, $^{2}$Laser Physics Centre, and
$^{3}$Centre for Ultrahigh-bandwidth Devices for Optical Systems
(CUDOS), Research School of Physical Sciences and Engineering,
Australian National University, Canberra, ACT 0200, Australia}
\vspace{-3mm}
\homepage{http://www.rsphysse.anu.edu.au/nonlinear}

\begin{abstract}
We study nonlinear coupling of mutually incoherent beams associated with different Floquet-Bloch waves in a one-dimensional optically-induced photonic lattice. We demonstrate experimentally how such interactions lead to asymmetric mutual focusing and, for waves with opposite diffraction properties, to simultaneous focusing and defocusing as well as discreteness-induced beam localization and reshaping effects.
\end{abstract}




\section{Introduction}
Light propagation in nonlinear media is associated with many important phenomena including optical rectification, harmonic generation, self-focusing, and soliton formation. Coupling of two or more optical waves mediated by a nonlinear material is responsible for four-wave mixing, parametric generation and amplification, as well as cross phase modulation and cross coupling. {\em Mutual beam focusing} involves both self-focusing and cross phase modulation of co-propagating beams and is an important effect with potential for all-optical light control and switching applications. Mutual beam focusing is also a fundamental phenomenon responsible for the formation of vector optical solitons in homogeneous dielectric media~\cite{Kivshar:2003:OpticalSolitons}.

Periodicity of the medium adds to the diversity of effects associated with nonlinear light propagation. In a material with a periodic modulation of the refractive index, the linear transmission spectrum becomes subject to dramatic modifications such as for instance the appearence of bandgaps, and the light propagation displays many new features that are governed by the properties of the extended eigenmodes of the periodic structure, the Floquet-Bloch waves. For example, beams associated with different Floquet-Bloch modes possess different diffraction properties, and this leads to normal, zero, and even anomalous diffraction in the same material~\cite{Eisenberg:2000-1863:PRL}. Furthermore, the bandgap spectrum of a periodic medium strongly affects nonlinear propagation and localization of light in the form of spatial solitons~\cite{Kivshar:2003:OpticalSolitons}. Different types of such self-trapped optical wavepackets have been demonstrated experimentally in nonlinear periodic structures, including discrete (or lattice) solitons in the semi-infinite total internal reflection gap~\cite{Eisenberg:1998-3383:PRL, Fleischer:2003-23902:PRL, Neshev:2003-710:OL}, and spatial gap solitons in the Bragg reflection gaps~\cite{Fleischer:2003-23902:PRL, Fleischer:2003-147:NAT, Mandelik:2004-93904:PRL, Neshev:2004-83905:PRL}. Vector solitons consisting of mutually coupled components localized in different gaps have been predicted theoretically~\cite{Cohen:2003-113901:PRL, Sukhorukov:2003-113902:PRL, Buljan:2004-223901:PRL, Motzek:2005-2916:OE}. Most recently, it was demonstrated experimentally that partially coherent beams can excite modes in several gaps that experience mutual trapping in media with slow photorefractive nonlinearity~\cite{Cohen:2005-500:NAT}. Mutual focusing of orthogonally polarized waves excited in different bands, as well as formation of single-polarization multi-band breathers, has been observed in AlGaAs waveguide array structures~\cite{Mandelik:2003-253902:PRL, Lahini:2005-1762:OE}. 

In this work, we study experimentally nonlinear interactions of mutually incoherent beams associated with different Floquet-Bloch modes of an optically-induced photonic lattice with self-focusing nonlinearity. The setup was designed to allow for simultaneous excitation of waves associated with the edges of the first and second transmission bands, and for independent control over the beam widths and amplitudes. Such flexibility enabled us to reveal novel effects for nonlinear beam propagation regimes that were not accessed in other experimental studies of interband mutual focusing~\cite{Cohen:2005-500:NAT, Mandelik:2003-253902:PRL, Lahini:2005-1762:OE}.
In particular, we demonstrate an inherent modal asymmetry in the effect of discrete mutual focusing of beams generated at the upper edges of the first two transmission bands. We observe transitional mutual focusing and defocusing of waves exhibiting diffraction of different magnitude and sign, and study in detail discrete beam break-up induced by highly localized fields. The results allow for generalization of symbiotic type interactions in periodic media, as predicted for pulses with different dispersion in optical fibres~\cite{Trillo:1988-871:OL, Afanasyev:1989-805:OL, Kivshar:1991-195:PS}

We consider a one-dimensional lattice induced optically in a
photorefractive crystal with strong electro-optic anisotropy, as
first suggested theoretically by Efremidis {\em et al.}~\cite{Efremidis:2002-46602:PRE}. Propagation of several mutually incoherent beams in such a lattice is described by a
system of coupled nonlinear Schr\"odinger equations for the normalized beam
envelopes $E_n(x,z)$,
\begin{equation} \leqt{nls}
   i \frac{\partial E_n}{\partial z}
   + D \frac{\partial^2 E_n}{\partial x^2}
   + {\cal F}( x, I) E_n
   = 0 ,
\end{equation}
where $I = \sum_n |E_n|^2$ is the total intensity, $x$ and $z$ are
the transverse and propagation coordinates normalized to the
characteristic values $x_0$ and $z_0$, respectively, $D = z_0
\lambda_0 / (4 \pi n_0 x_0^2)$ is the beam diffraction coefficient,
$n_0$ is the average refractive index of the medium, and $\lambda_0$
is the wavelength in vacuum. The optically-induced refractive
index change is ${\cal F}( x, |E|^2) = - \gamma (I_b + I_p(x) + |E|^2)^{-1}$,
where $I_b$ is the constant dark irradiance, $I_p(x)=I_g
\cos^2(\pi x / d)$ is the interference pattern which induces a
lattice with period $d$, and $\gamma$ is a nonlinear coefficient
proportional to the applied DC field~\cite{Fleischer:2003-23902:PRL,Neshev:2003-710:OL}. To match our experimental
conditions, we use the following parameters: $\lambda_0 = 0.532~\mu$m, $n_0 = 2.35$, $x_0 = 1~\mu$m, $z_0 = 1$mm, $d = 19.2$, $I_b =1$ and $I_g =1$. Then, the refractive index contrast in the lattice is
$\Delta n = \gamma \lambda / (4 \pi z_0)$, which evaluates to $1\cdot10^{-4}$ for $\gamma = 2.36$. The crystal length is $L = 15$~mm.

\begin{figure}[H]
	\centerline{\vspace{-3mm}
  \includegraphics[width=0.95\columnwidth]{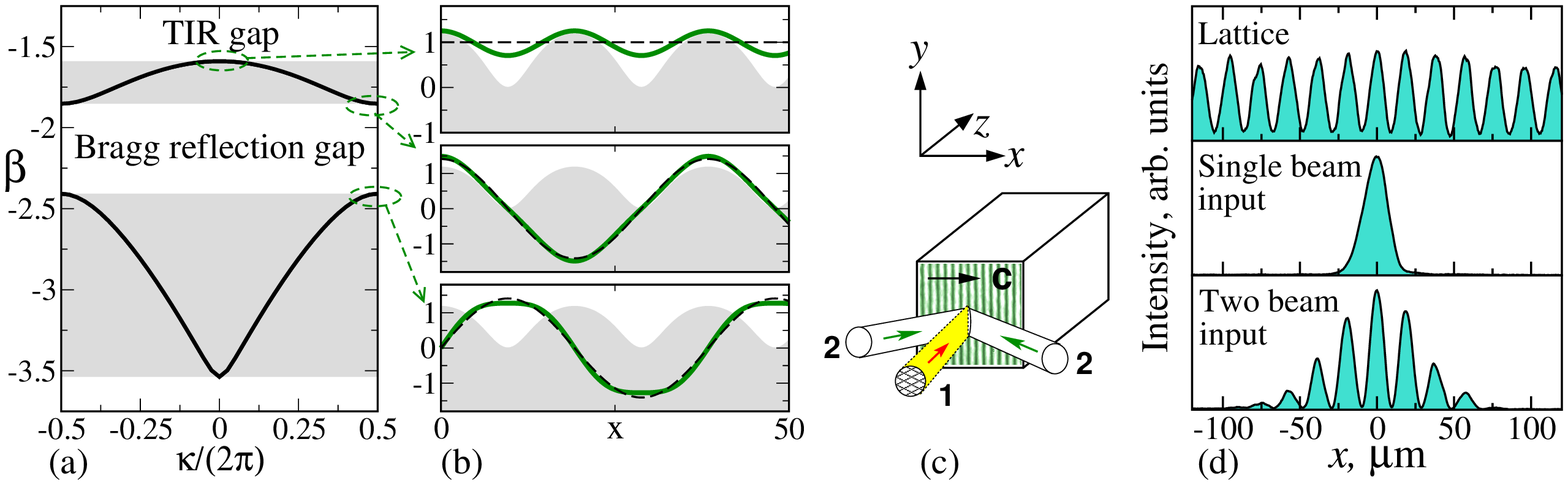}}
  \protect\caption{\protect\label{fig:scheme} (a)~Dispersion of the Bloch waves in an optically-induced lattice; the bands are shaded. (b)~Bloch-wave
profiles at the gap edges (solid) and their leading-order Fourier
components (dashed), superimposed on the normalized refractive
index profile (shaded). (c)~Schematic of (1) the Bloch-wave
excitation with a single beam and (2) a pair of beams inclined at the Bragg angle. (d)~Experimental intensity profiles of the optical lattice
and the input beams.}
\end{figure}

\section{Experimental setup}
We create an optical lattice with a period of
19.2~$\mu$m inside a $15\times5\times5$~mm SBN:60 crystal by
interfering two ordinarily-polarized broad laser beams from a
frequency-doubled Nd:YVO$_4$ laser at 532~nm. Applying an external
electric field of $1.8-2.0$~kV then produces a periodic refractive
index modulation whose saturation is controlled by homogeneously
illuminating the crystal with white light. The induced lattice
gives rise to a linear transmission spectrum with bandgaps as illustrated in
Fig.~\rpict{scheme}(a). 
Extraordinarily-polarized probe beams from the same laser source are focused onto the front
face of the crystal and launched into the lattice in a way as to selectively excite Bloch waves associated with the edges of the bandgaps.
The eigenmode profiles at the gap edges
are shown in Fig.~\rpict{scheme}(b) along with the corresponding
refractive index modulation (shaded). 

In order to excite the specific Bloch waves, we use a multiple-beam technique,
schematically shown in Fig.~\rpict{scheme}(c). To obtain pure
excitation of a given Bloch wave, the input beam must approximate
the eigenmode profile by matching its leading-order Fourier
component [Fig.~\rpict{scheme}(b-dashed line)]. Hence, a single
Gaussian beam [Fig.~\rpict{scheme}(d-middle)] is used to excite the fundamental wave at the top of
the first band (i.e. at the edge of the total internal reflection gap)
~\cite{Neshev:2003-710:OL}, and two overlapping coherent
beams, both inclined at the Bragg angle, are used to excite Bloch
waves with a staggered phase structure at the two edges of the Bragg reflection
gap~\cite{Neshev:2004-83905:PRL}. Aligning the maxima of the
lattice and the two-beam interference profile [Fig.~\rpict{scheme}(d-bottom)], we obtain pure excitation of the
Bloch mode at the bottom of the first band. Shifting the probe
beam position by half a period leads to a pure excitation of the
mode at the top of the second band. The beams {\sf 1} and {\sf 2}
in Fig.~\rpict{scheme}(c) are made mutually incoherent by
reflecting beam {\sf 1} from a piezo-transducer controlled mirror
vibrating at high frequency. All excited Bloch waves propagate straight
down the lattice with zero transverse velocity, in case of the staggered modes thanks to an exact power balance between the two inclined input beams. 
At the crystal output, the
near-field intensity distribution of the propagated beams is
imaged onto a CCD camera.

\section{Discrete interband mutual focusing}
We first study discrete {\em interband} mutual coupling of two
beams associated with the top of the first and the second band.
Both exhibit {\em normal} diffraction and can therefore be
localized at higher powers by the self-focusing nonlinearity. This leads to the effect of mutual focusing since each beam is capable of inducing a local defect (waveguide) that traps the other component.
Mutual focusing can occur for a wide range of different relative
and absolute beam powers. Here we investigate the extreme cases 
in which the beam powers are
strongly imbalanced, and where the coupling effects are therefore
most easily observed. We initially set the power of both beams at
a low (nW) level so that they propagate in a linear regime. Then the
power of one of the components is increased and the nonlinear
effect on both beams is studied. Unless otherwise stated, the two probe beams have an input
full width at half maximum (FWHM) of $17\pm3~\mu$m and
$70\pm3~\mu$m, respectively.

\begin{figure}[H]
	\centerline{\vspace{-3mm}
  \includegraphics[width=0.8\columnwidth]{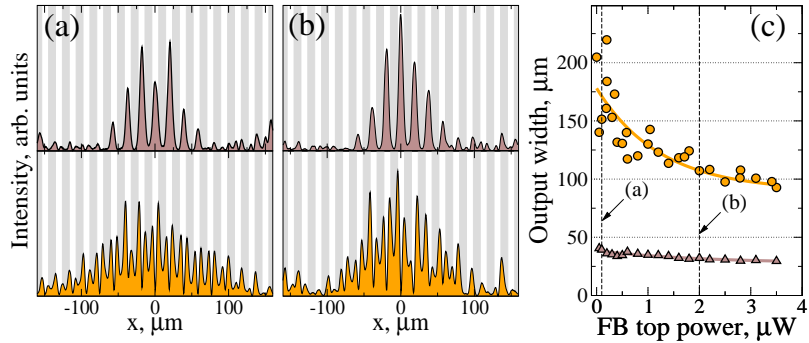}}
  \protect\caption{\protect\label{fig:fig1} Experimental results for interband
mutual focusing of the beams generated at the top of the $2^{nd}$ band
(orange) and top of the $1^{st}$ band (brown). (a) Both components
at low power. (b) $1^{st}$-band beam at high power and
$2^{nd}$-band beam at low power ($70$~nW). (c) Beam width vs.
power of the $1^{st}$-band mode. Solid curves are exponential fits
to the data points.}
\end{figure}

Figure~\rpict{fig1} shows the result
of increasing the power of the fundamental first band beam
(brown). Transition from discrete diffraction of this beam in the
linear regime [Fig.~\rpict{fig1}(a)] to discrete focusing at higher power [Fig.~\rpict{fig1}(b)] is
accompanied by a decrease in the FWHM of the broader
co-propagating second-band beam by almost a factor of two. The
second-band beam width is estimated by a Gaussian envelope fitting, whereas
the width of the first-band beam is described by its discrete
second-order central moment
$w^2=\sum2^{\left|n\right|}d^2P(n)/\sum{P(n)}$, where $2n+1$ is
the number of spanned lattice sites, $d$ is the lattice period,
and $P(n)$ is the intensity of the $n^{th}$ site. We calculate
$w^2$ for the three central lattice sites ($n=1$) and characterize
the beam width by $2w$. To investigate the effect in further
detail, we measure the degree of focusing of both components as a
function of the first-band beam power. The result is shown in
Fig.~\rpict{fig1}(c), and we notice that the width of the passive
second-band beam asymptotically shrinks from $\sim180\mu$m to
$\sim90\mu$m, whereas the effect is rather small for the
first-band beam itself.

\begin{figure}[H]
	\centerline{\vspace{-3mm}
  \includegraphics[width=0.85\columnwidth]{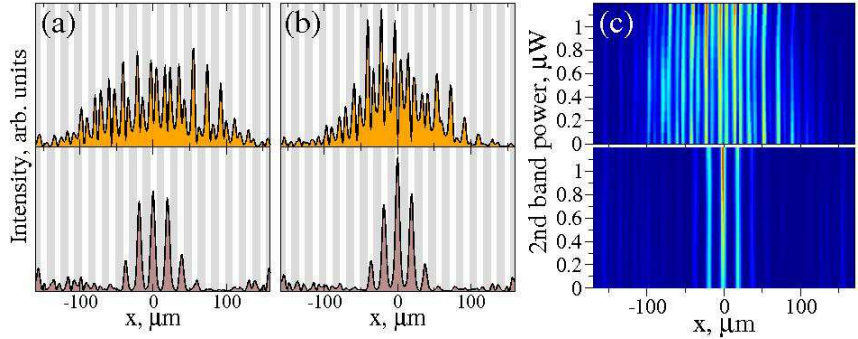}}
  \protect\caption{\protect\label{fig:fig2} Experimental results for interband
mutual focusing of the beams generated at the top of the $1^{st}$
band (brown) and top of the $2^{nd}$ band (orange). (a) Both
components at low power. (b) $2^{nd}$-band beam at high power
($900$~nW) and $1^{st}$-band beam at low power ($100$~nW). (c)
Two-dimensional visualization of output profiles for increasing
$2^{nd}$-band power.}
\end{figure}

Simultaneous beam focusing is also observed if, conversely, the
power of the first-band component is maintained at a low level and
that of the second band beam is increased [Fig.~\rpict{fig2}].
Since the defect induced by the high-power second-band beam is
rather broad and weak, we observe less pronounced focusing of the
first-band beam in this case. In fact, it was necessary to slightly increase
the input beam width to $20~\mu$m in order to reduce diffraction
and thereby facilitate focusing, and that is why no clear discrete
diffraction is observed for linear propagation of the first-band beam in Fig.~\rpict{fig2}(a). 
Figure \rpict{fig2}(c) maps the beam profiles as a function of the second-band beam power in a two-dimensional plot.

Together the above results demonstrate discrete interband mutual focusing of mutually incoherent beams, an effect responsible for the formation of multi-gap vector solitons. The observed asymmetry in the strength of the coupling between the bands reflects the fact that the physics of the discrete total internal reflection and Bragg gap solitons associated with the involved Bloch waves is quite different. Indeed, a discrete soliton in the total internal reflection gap creates a narrow and strongly localized defect, whereas both the minimum width and the maximum intensity of a Bragg gap soliton are limited due to the finite spectral width of the Bragg reflection gap. This means that there is also a limit for the trapping strength of the self-induced second-band defect. The above-mentioned generic features may well account for the saturation of the second band mode focusing [Fig.~\rpict{fig1}(c)], and the observed interband coupling asymmetry, i.e. the fact that the two beams are not equally capable of trapping each other.

\section{Interaction of beams with normal and anomalous diffraction}
Next we study {\em intraband} coupling effects for the two modes at
the top and the bottom of the first band. These waves exhibit
{\em normal} and {\em anomalous} diffraction, respectively, and therefore the first self-focuses
while the second experiences defocusing in the self-focusing nonlinearity. However, the picture is more
complicated when the two waves interact via the nonlinear medium. We show that in this case
mutual coupling can lead to simultaneous focusing and defocusing
and to interesting discrete localization effects. Indeed, when the
power of the beam associated with the bottom of the band is
increased [Fig.~\rpict{fig3}] it does spread out stronger due to
self-defocusing (blue). However, it still induces a
lattice defect that causes the mode at the top of the band (brown)
to focus. This is to our knowledge a first demonstration of simultaneous mutual
focusing and defocusing of beams belonging to different parts of
the bandgap spectrum. Figure~\rpict{fig3}(c) shows that the beam widths
increase and decrease almost linearly as a function of the power of the beam at the top of the band.
We would like to
point out that whereas the interband mutual focusing can lead to
the formation of stable multi-gap vector solitons, the intra-band
coupling effects responsible for the observed simultaneous
focusing and defocusing [Fig.~\rpict{fig3}] are inherently {\em
transient} in nature since at longer propagation distances, the
two beams will eventually decouple from each other. However, soliton formation is not in general required for obtaining e.g. all-optical switching, and even spatially transient effects may be sufficient for achieving effective beam manipulation functionalities. 

In the opposite case [see Fig.~\rpict{fig4}], the beam at the top
of the band self-focuses at high powers, but induces a repelling
defect for the defocusing beam at the bottom of the band. We
observe that as the power of the fundamental beam is increased
[movie in Fig.~\rpict{fig4}(c)] the power carried by the two
lattice sites (index maxima) next to the central one
drops significantly at the crystal output. Eventually, at high
power, the output profile is composed of a single centrally
localized peak and two discretely repelled outer lobes. Clearly, this complex beam reshaping cannot be described by a simple envelope approximation and calls for further investigations.

\begin{figure}[H]
	\centerline{\vspace{-3mm}
  \includegraphics[width=0.8\columnwidth]{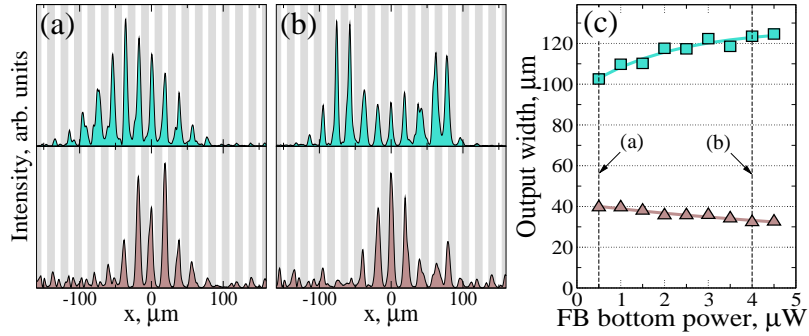}}
  \protect\caption{\protect\label{fig:fig3} Experimental results for coupling of beams at the top (brown)
and bottom (blue) of the $1^{st}$ band. (a) Both components at low
power. (b) Bottom of $1^{st}$ band at high power and top of
$1^{st}$ band at low power ($100$~nW). (c) Beam width vs. power
of the $1^{st}$ band bottom mode.}
\end{figure}

\begin{figure}[H]
	\centerline{\vspace{-3mm}
  \includegraphics[width=0.85\columnwidth]{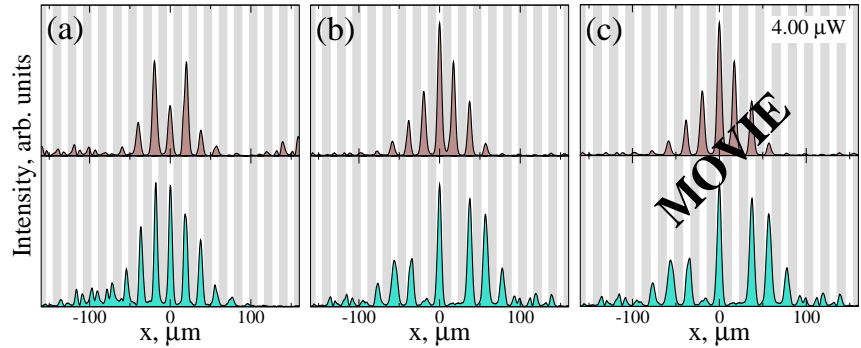}}
  \protect\caption{\protect\label{fig:fig4} Experimental results for coupling of beams at the bottom
(blue) and top (brown) of the $1^{st}$ band. (a) Both components
at low power ($50$~nW). (b) Top of $1^{st}$ band at high power and bottom of
$1^{st}$ band at low power. (c) Movie (0.4 Mb) shows the
fundamental beam self-focusing and the simultaneous appearance of
a discrete localized peak and two decoupled outer lobes for the
beam at the bottom of the band. The power of the fundamental beam
is shown in the upper right corner.}
\end{figure}

\section{Numerical simulations}
To support our experimental results and to more closely examine the
features associated with discrete localization and beam shaping effects for waves with opposite diffraction properties, a series of extended numerical simulations were performed. They confirmed all important aspects of our experimental observations, including the asymmetric interband coupling, the transitional mutual focusing and defocusing, and the discreteness-induced reshaping of the anomalously diffracting first-band beam. Fig.~\rpict{fig5}
shows two examples of simulations representing the experimental cases of mutual interband focusing in Fig.~\rpict{fig1} and first band focusing-defocusing interactions in Fig.~\rpict{fig4}. The simulated propagation distance corresponds to the length of our crystal.

\begin{figure}[H]
	\centerline{\vspace{-3mm}
  \includegraphics[width=0.95\columnwidth]{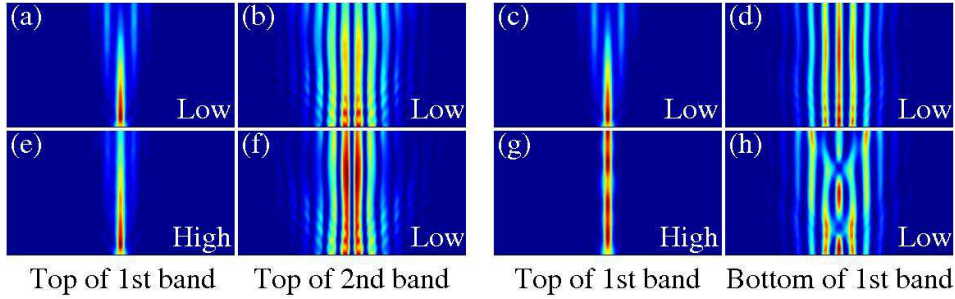}}
  \protect\caption{\protect\label{fig:fig5} Beam propagation method simulations of interband
focusing (left) and interaction of focusing and defocusing
$1^{st}$-band beams (right). The plot windows are $320~\mu$m and
$15~$mm in the horizontal (transverse) and vertical (propagation) directions,
respectively. In (a,b) and (c,d) two beams co-propagate at low
power and experience diffraction. In (e,f) the power of the beam
at the top of the $1^{st}$ band is increased, causing focusing of
this beam and the low power $2^{nd}$-band beam. In (g,h)
interaction between a high power fundamental beam and a low power
beam at the bottom of the $1^{st}$ band results in the formation
of a complex beating pattern and the appearance of a
localized discrete central peak and two diffracting outer lobes.}
\end{figure}

The first case
[Fig.~\rpict{fig5},~left] presents a pair consisting of a fundamental and a
second band beam which both experience diffraction broadening
at low power (top panels) and then focus as the power of the first band
component is increased (bottom panels). Figure~\rpict{fig5}~(right) shows a pair of beams with opposite
dispersion from the top and bottom of the first band. At low
power, both components propagate linearly and diffract. At high
power, the fundamental mode (top of the band) self-focuses and
induces a local lattice defect, which results in a complex
propagation pattern for the anomalously diffracting wave at the
bottom of the band. First, power seems to oscillate back and forth
between the central lattice site and its two adjacent neighbours,
indicating a beating between modes with different propagation
constants. Then, a central peak and two clearly decoupled outer
lobes eventually appear, as observed experimentally. Simulations
for extended propagation distances (not shown) revealed that after
about 50\% longer propagation the oscillatory behavior ceases, and
the decoupled power propagates away from the defect in two
distinct lobes at the zero diffraction angles. Interestingly, a
considerable part of the power stays trapped in the induced defect
and propagates straight down the lattice as a stable waveguide
mode. In all simulations the FWHM of the fundamental and the Bragg
input beams was $15~\mu$m and $70~\mu$m, respectively, and the
peak power of the fundamental first-band beam (normalized to a constant
dark irradiance $I_b$) was $0.25$ and $3.24$ in
Figs.~\rpict{fig5}~(e,f) and (g,h), respectively.

\section{Conclusions}
In conclusion, we have studied interaction of mutually incoherent
beams associated with Floquet-Bloch waves from different
spectral bands. We have demonstrated experimentally how defects induced by nonlinear propagation can trap and guide beams
from several bands, leading to the effect of discrete
interband mutual focusing. This is a key physical mechanism for
generation of multi-gap solitons with components originating
from different spectral bands. We have observed a fundamental asymmetry in the interband mutual focusing effect due to the generically different physics of the involved Bloch waves. We also demonstrated an interplay
between waves with diffraction coefficients of opposite sign that leads
to generation of anti-waveguiding defects, complex beam reshaping and simultaneous
focusing and defocusing within the first transmission band.

\end{sloppy}
\end{document}